\documentclass[11pt]{article}
\usepackage{latexsym}  
\usepackage{amssymb}
\usepackage{graphicx}
\usepackage{amsmath}

\newcommand{\be}{\begin{equation}}
\newcommand{\ee}{\end{equation}}
\newcommand{\bea}{\begin{eqnarray}}
\newcommand{\eea}{\end{eqnarray}}
\newcommand{\bref}[1]{(\ref{#1})}

\newcommand{\bi}{\bibitem}
\newcommand{\pa}{\partial}

\usepackage[bottom=30mm]{geometry}
\begin{document}
\begin{titlepage}
\begin{flushright}
\today
\end{flushright}
\vspace{4\baselineskip}
\begin{center}
{\Large\bf Bose-Einstein Condensate Cosmology within the Framework of QCD Axion}
\end{center}
\vspace{1cm}
\begin{center}
{\large Takeshi Fukuyama
\footnote{E-mail:fukuyama@se.ritsumei.ac.jp}}
\end{center}
\vspace{0.2cm}
\begin{center}
{\small \it Research Center for Nuclear Physics (RCNP),
Osaka University, Ibaraki, Osaka, 567-0047, Japan}\\[.2cm]



\vskip 10mm
\end{center}
\vskip 10mm
\begin{abstract}
Bose–Einstein Condensation (BEC) cosmology is analyzed in the framework of a string-inspired axion model. The dispersion relation of the axionic mode includes both gravitational and self-interaction terms, the latter being small in magnitude but crucial for inducing instability of the condensate. The generation rate of BEC around redshift $z\approx 30$ is primarily governed by gravity, consistent with a phenomenological value $\Gamma\approx 10^{-31}$ eV adopted in previous work and realizable for the QCD axion. The relation between this early BEC epoch and the later formation of supermassive black holes at $z\approx 5$ is also discussed.

\end{abstract}
\end{titlepage}
Over the past decade, we have developed a cosmological model based on a boson gas and its unstable Bose–Einstein condensate (BEC), referred to as BEC Cosmology \cite{FM1, FM2, FM3}. Within this framework, a light (pseudo) scalar dark matter (DM) component can effectively behave as dark energy (DE) once it forms a BEC. In particular, we focus on scalar particles with an attractive quartic self-interaction, which renders the condensate unstable, as will be discussed below.
This model was motivated by the aim to address several key issues:

i) a novel mechanism for cosmic inflation,
ii) a natural explanation of the “why now” problem (the coincidence between the present magnitudes of DM and DE),
iii) the very early formation of black holes (BHs), and
iv) the logarithmic–redshift periodicity observed in the mass density of the universe \cite{Broadhurst, DESI}.

Under the assumption that DM exhibits self-attraction rather than self-repulsion, the resulting BEC becomes inherently unstable. The evolution of the system is then governed by the following set of equations \cite{FM1}.
\begin{eqnarray}
H^{2} &=&\left( {\frac{\dot{a}}{a}}\right) ^{2}=\frac{8\pi G}{3c^{2}}\left( {%
\rho _{g}+\rho _{\phi }+\rho _{l}}\right) \;,  \nonumber \\
\dot{\rho}_{g} &=&-3H\rho _{g}-\Gamma \rho _{g} \;,  \nonumber \\
\dot{\rho}_{\phi } &=&-6H\left( {\rho _{\phi }-V}\right) +\Gamma \rho _{g}-{%
\Gamma }^{\prime }\rho _{\phi } \;,  \nonumber \\
\dot{\rho}_{l} &=&-3H\rho _{l}+{\Gamma }^{\prime }\rho _{\phi } \;. 
\label{eq16}
\end{eqnarray}%
Here, $a$ denotes the scale factor (remark that, hereafter, the subscript $a$ refers to the axion). The quantities $\rho_g,~\rho_\phi$, and $\rho_l$ represent the energy densities of the DM gas, the DM condensate, and the localized DM produced by the decay of the condensate, respectively. The parameter $\Gamma$ denotes the growth rate of the boson gas into the BEC phase, while $\Gamma^\prime$ represents the decay rate of the BEC into a collapsed condensate (i.e., localized DM). The former, $\Gamma$, is treated as a constant, whereas $\Gamma^\prime$ becomes nonzero only when the BEC satisfies the instability condition. These rates act as transport coefficients characterizing the BEC phase transition. Their values were phenomenologically determined in Refs.~\cite{FM1, FM2} and, for example we set the maximal growth rate of BEC are set as
\be
\Gamma=1\times 10^{-31}~~\mbox{eV}
\label{G1}
\ee
by hand. 
So far, we have not associated this scalar particle with any known model. Interestingly, however, the QCD axion \cite{PQ1, PQ2, KSVZ1, KSVZ2, ZDFS1, ZDFS2}—more specifically, a string-inspired axion \cite{Svrcek}—presents a compelling candidate that naturally fulfills the role of  our scalar field, and we will provide the expression for $\Gamma$ in Eq. \bref{Gamma}.

We start with the axion Lagrangian,
\be
\mathcal{L}=\frac{1}{2}\pa_\mu\phi\pa^\mu\phi-V_a(\phi)
\label{ReL}
\ee
with axion potential 
\bea 
V_a&=&\Lambda^4\left(1-\cos \left(\frac{\phi}{f_a}\right)\right)\\
&=&m_a^2f_a^2\left(1-\cos \left(\frac{\phi}{f_a}\right)\right)\approx \frac{1}{2}m_a^2\phi^2+\frac{1}{4!}\lambda \phi^4 \;,
\eea
where 
\be
\lambda\equiv -\frac{m_a^2}{f_a^2} \;.
\ee
The 2 → 2 scattering cross section for this particle is
\be
\sigma=\frac{\lambda^2}{64\pi m_a^2}\equiv 4\pi s_l^2 \; ,
\ee
where $s_l$ is the scattering length, defined by
\be
s_l=\frac{\lambda}{16\pi m} \;.
\label{slength}
\ee
Axion field potential in the classic dilute gas approximation becomes
\be
V(\phi)=m_u\Lambda_{QCD}^3\left[1-\cos\left(\frac{\phi}{f_a}\right)\right] \;.
\label{potential}
\ee
Using the Gellmann-Oakes-Renner relation \cite{Gellmann}, 
\be
\Lambda_{QCD}^3=\frac{F_\pi^2m_\pi^2}{m_u+m_d} \;.
\label{LQCD}
\ee
Here $F_\pi=93$ MeV, and $\frac{m_u}{m_d}\approx 0.47$, and we obtain axion mass
\be
m_a=5.7\times 10^{-6}\left(\frac{10^{12}\mbox{GeV}}{f_a}\right) \text{eV} \;.
\label{ma}
\ee
Then, the axion self coupling constant $\lambda$ in $\frac{\lambda}{4!}\phi^4$ becomes
\be
\lambda=-0.47\frac{F_\pi^2m_\pi^2}{f_a^4}<0 \;.
\label{selfc}
\ee
On the other hand, heterotic superstring theory gives on the same QCD axion the different form, 
\be
\Lambda^4\rightarrow \Lambda_{string}^4=M_{SUSY}^2M_{Pl}^{*2}e^{-S_{instanton}} 
\label{Lstring}
\ee
and $f_a$ is given by \cite{Svrcek}
\be
f_a=\frac{\alpha_{GUT}M_{Pl}^*}{\sqrt{2}2\pi}\approx 1.1\times 10^{16}~~\mbox{GeV} \;.
\label{fa}
\ee
Here we have set the unified coupling constant 
\be
\alpha_{GUT}=\frac{1}{25} \;.
\ee
As for the axion mass $m_a$, it depends critically on the instanton action $S_{instanton}$ \cite{Svrcek, Visinelli, Fuku1}, yielding an estimated value of
\begin{equation}
m_a \simeq 1 \times 10^{-22}~\text{eV} \;.
\end{equation}
It is noteworthy that these two types of axions ($a_i$) can coexist.

Write two canonically normalizable axions $a_1,~a_2$ with two decay constant $f_1,~f_2$ and QCD anomaly coefficients $c_1,~c_2$. Their QCD coupling can be written as
\be
\mathcal{L}_{QCD}=\frac{\alpha_s}{8\pi}(q^Ta)G\tilde{G},~~a\equiv (a_1, a_2)^T \;, ~~q\equiv \left(\frac{c_1}{f_1},\frac{c_2}{f_2}\right)^T \;.
\ee
The QCD potential selects the direction parallel to $q$. The physical QCD axion is therefore
\be
a_{QCD}=\frac{q^Ta}{||q||},~~||q||=\sqrt{\frac{c_1^2}{f_1^2}+\frac{c_2^2}{f_2^2}} 
\ee
with effective decay constant $f_{eff}=1/||q||$ and mass $m_{aQCD}\approx 5.7\mu\mbox{eV}(10^{12}\mbox{GeV}/f_{eff})$. The orthognal combination
\be
a_\perp=\frac{1}{||q||}\left(-\frac{c_2}{f_2},\frac{c_1}{f_1}\right)\cdot a
\ee
does not couple to $G\tilde{G}$ at leading order and is an ALP; its mass must arise from non-QCD dynamics (e.g. hidden instantons) the simplest case, its effective decay constant is
\be
\frac{1}{f_{eff}^2}=\frac{c_1^2}{f_1^2}+\frac{c_2^2}{f_2^2} \;.
\ee
If $f_2\gg f_1$ and $c_1,~c_2\approx O(1)$, then $f_{eff}\approx f_1$. So the QCD-solving model is dominantly $a_1$, while $a_2$ behaves as an ultra-weakly coupled ALP. 
Here we are considering cosmology and the string-inspired axion plays the main role as we will discuss.

First, it should be noted that the situations are quite different in the relativistic and nonrelativistic cases. In the relativistic case, the field $\phi$ must undergo quantum tunneling toward the stable point, and the decay rate is therefore strongly suppressed. However, if a nonrelativistic treatment is adopted, the situation may change significantly. The transition from the relativistic to the nonrelativistic regime is achieved by rescaling the field as
\be
\phi=\frac{1}{\sqrt{2m}}\left(\psi e^{-im_at}+\psi^*e^{im_at}\right)=\phi^* \;,
\ee
where $\psi$ represents the slowly varying condensate wavefunction. 
Non-relativistic version is
\be
i\pa\psi/\pa t=-(1/2m_a)\nabla^2\psi+V\psi+g|\psi|^2\psi \;,
\ee
where $V(x)$ is the potential, and
\be
\lambda=8g m_a^2 \;.
\label{lambda}
\ee
The dispersion relation gives
\be
\omega_k^2=(k^2/2m_a)^2-4\pi G\rho+(k^2/2m_a)\times 2gn
\label{omega}
\ee
with axion number density $n$.
In this paper we consider the string-inspired axion:
\be
m_a=10^{-22}~\mbox{eV} \;, ~~f_a=10^{16}~\mbox{GeV} \;, ~~g=-f_a^{-2} \;.
\ee
In this case the contact collisions are negligible ($\sigma\propto m_a^2/f_a^4$). The rate is controlled by gravity:
\be
\Gamma_{grav}\approx \sqrt{4\pi G\rho} \;,~~t_{grow}\approx (4\pi G\rho)^{-1/2} \;.
\label{Gamma}
\ee
We set in Ref. \cite{FM2} 
\be
\Gamma=1\times 10^{-31}~\mbox{eV},~~\mbox{and}~~t_{grow}\approx 2\times 10^8~\mbox{yrs} \;,
\ee
which effectively matched with $\sqrt{4\pi G\rho}$ at $z=30$.
On the other hand, in our previous paper \cite{Fuku1}, we argued this ultra-light axion in Super Massive Black Holes (SMBHs). In that case $\Gamma_n=|g|n$ dominates.
A useful diagnostics is
\be
S\equiv \frac{|g|n}{\sqrt{4\pi G\rho}}=\frac{|g|}{\sqrt{4\pi G\rho m_a}}, ~~n_{crit}=\frac{4\pi Gm_a}{|g|^2} \;.
\ee
If $S\ll 1$ (or $n\ll n_{crit}$): self-attraction negligible $\rightarrow$gravity +quantum pressure dominate.
If $S\geq 1$ (or $n\geq n_{crit}$): attractive nonlineality assists collapse (SMBH case).
\begin{table}[h!]
\centering
\caption{Cosmic mean dark-matter quantities for $m_a=10^{-22}\,\mathrm{eV}$, $f_a=10^{16}\,\mathrm{GeV}$.}
\vspace{0.5em}
\begin{tabular}{cccccc}
\hline
$z$ & $\rho_{\rm DM}\,[{\rm GeV/cm^3}]$ & $n\,[{\rm cm^{-3}}]$ &
$|g|n\,[{\rm eV}]$ & $\sqrt{4\pi G\rho}\,[{\rm eV}]$ \\
\hline
5  & $1.5\times10^{-6}$ & $1.5\times10^{25}$ & $6.2\times10^{-38}$ & $1.3\times10^{-32}$ \\
10 & $1.2\times10^{-5}$ & $1.2\times10^{26}$ & $5.2\times10^{-37}$ & $4.4\times10^{-32}$ \\
30 & $3.9\times10^{-4}$ & $3.9\times10^{28}$ & $1.6\times10^{-35}$ & $1.1\times10^{-31}$ \\
\hline
\end{tabular}
\label{tab:cosmic_n_gG}
\end{table}
Although the quartic coupling $g=-1/f_a^2$ is extremely small in magnitude for $f_a=10^{16}$ GeV, its negative sign has a crucial qualitative effect in the Gross-Pitaevskii–Poisson system \cite{GP}. In the homogeneous background, $|g|n\ll \sqrt{4\pi G\rho}$, so the self-interaction does not set the condensation rate. Nevertheless, the attractive sign renders the BEC marginally unstable, allowing part of the condensate to collapse while another part decays back to the non-condensed gas, as described in Ref. \cite{FM2}. Thus, the self-attraction is dynamically essential though numerically subdominant.
Although the BEC transition discussed in Ref. \cite{FM2} occurs at an earlier cosmic epoch ($z\approx 30$) than the supermassive–black–hole formation discussed in Ref. \cite{Fuku1}, the corresponding number densities differ because the two models describe distinct physical environments. Ref. \cite{FM2} treated the homogeneous background conversion of dark–matter gas to a cosmic BEC at the mean density $n\approx (10^{28}-10^{29})/\mbox{cm}^3$, whereas the later SMBH work concerns locally collapsed BEC regions whose densities reach $n_{halo}\approx(10^{33}-10^{35})/\mbox{cm}^3$. As structure formation progresses, the mean density decreases as $(1+z)^3$ but gravitational clustering generates overdensities of many orders of magnitude. Hence the later event naturally exhibits higher local density despite occurring at a smaller redshift; the two results are therefore fully compatible.

\section*{Acknowledgements}

This work is supported in part by the Grant-in-Aid for Science
Research from the Ministry of Education, Science and Culture of Japan
(No. 21104004).

\end{document}